\documentstyle[12pt,epsf,rotate]{article}

\newcommand{\be}{\begin{equation}}
\newcommand{\ee}{\end{equation}}
\newcommand{\bea}{\begin{eqnarray*}}
\newcommand{\eea}{\end{eqnarray*}}
\newcommand{\ba}{\begin{eqnarray}}
\newcommand{\ea}{\end{eqnarray}}
\newcommand{\MSbar}{\overline{\rm MS}}
\newcommand{\lamFP}{\lambda_{\rm FP}}

\begin{document}

\begin{flushright}
August 1997\hfill DESY 97-152\\
DO-TH 97/18
\end{flushright}

\vspace*{1cm}
\begin{center}\large{\bf 
SM Higgs mass bounds from theory
\footnote{To appear in the Proceedings of the
``{\it ECFA/DESY Study on Physics and Detectors for the Linear
  Collider}'', ed. R. Settles, DESY publication DESY 97-123E.}
}\end{center}

\vspace*{2mm}

\begin{center}{Thomas Hambye$^{\dagger}$ and Kurt Riesselmann$^*$}
\end{center}
\vspace*{0mm}
\small

\vspace{-5mm}

\begin{center}
$^{\dagger}$ Institut f\"ur Physik T3, Universit\"at Dortmund, 
D- 44221 Dortmund, Germany
\end{center}
\vspace{-9mm}
\begin{center}
$^*$ DESY Zeuthen, Platanenallee 6, D-15738 Zeuthen, Germany
\end{center} 

\vspace*{2mm}

\small
\begin{abstract}
  The two-loop Higgs mass upper bounds are reanalyzed.  Previous
  results for a cutoff scale $\Lambda\approx$ few TeV are found to be
  too stringent.  For $\Lambda=10^{19}$ GeV we find $M_H < 180 \pm
  4\pm 5$ GeV, the first error indicating the theoretical uncertainty,
  the second error reflecting the experimental uncertainty due to
  $m_t=175\pm6$ GeV.  We also summarize the lower bounds on $M_H$. We
  find that a SM Higgs mass in the range of 160 to 170 GeV will
  certainly allow for a perturbative and well-behaved SM up to the
  Planck-mass scale $\Lambda_{Pl}\simeq 10^{19}$ GeV, 
  with no need for new physics to
  set in below this scale.
\end{abstract}
\normalsize
\vspace{8mm}

It is well known that the high energy behavior of the Standard Model
(SM) Higgs quartic coupling is unsatisfactory at high energies. For a
{\it heavy} Higgs boson it manifests itself in the (one-loop) Landau
pole~\cite{triviality} when using a perturbative approach, or in large
cutoff effects when performing lattice calculations~\cite{lattice,LW}.
If the Higgs boson is {\it light} the Higgs running coupling may
become negative at high energy, giving rise to the problem of vacuum
stability~\cite{stability}. Defining a cutoff scale $\Lambda$ both
problems are avoided if the Higgs boson mass $M_H$ is constrained from
below and above, with no need for introduction of physics beyond the
SM.

Previous work~\cite{oldvacuum,einhorn,oldwork,lindner} extensively
investigated the dependence of the $M_H$ bounds on the top quark mass
$m_t$.  By the time the next generation of colliders are operating,
the good experimental knowledge of $m_t$ will make the theoretical
bounds on the Higgs mass a single function of the cutoff scale
$\Lambda$.
However, there are theoretical uncertainties which remain in the
calculation of the $M_H$ bounds.  In the case of the $M_H$ {\it lower}
bound these uncertainties have been recently addressed in
Ref.~\cite{altisi,casas1,casas2}.  Here we review the sensitivity of
the {\it upper} bound on $M_H$ with regard to various cutoff criteria,
the inclusion of matching corrections, and the choice of the matching
scale $\mu_0$ using a two-loop perturbative approach~\cite{HR}. We
also summarize the recent results for the lower
bounds~\cite{altisi,casas1,casas2}.
If a SM Higgs boson is found, the future measurement of its mass can
immediately be used to determine up to which maximal energy scale the
Standard Model could be valid.

The high-energy evolution of the SM running couplings is determined by
the beta functions of the theory. The value of the Higgs and top quark
$\MSbar$ running couplings are fixed at low energies through the
matching conditions
\begin{eqnarray}
\label{mchiggs}
\bar \lambda (\mu_0) &=& \frac{M^2_H}{2v^2}  [1+\delta_H(\mu_0)]\,,\\
\label{mctop}
{\bar g_t (\mu_0)} &=& \frac{\sqrt{2} m_t}{v} [1+\delta_t(\mu_0)]\,,
\end{eqnarray}
where $v = (\sqrt{2} G_F)^{-1/2} \approx 246$ GeV, and $\mu_0$ is the
matching scale.  The definitions of the tree level couplings are
obtained by setting the matching corrections $\delta$ equal to zero,
thus fixing our notation.

In the present analysis, we restrict ourselves to two-loop beta
functions~\cite{einhorn,vaughn2} and the corresponding one-loop
matching conditions (see Ref.~\cite{sirlin} for $\delta_H$ and
Ref.~\cite{HK} for $\delta_t$) in the $\MSbar$ scheme.  For the
$\MSbar$ electroweak and strong couplings we take the input values
$\bar g(M_Z)=0.651$, $\bar g'(M_Z)=0.357$, and $\alpha_s(M_Z) =
0.118$.

For sufficiently large initial coupling $\lambda(\mu_0)$ the two-loop
running coupling $\lambda (\mu)$ approaches the (meta-stable)
fixed-point value $\lambda_{\rm FP} = 12.1...$ at some high energy
scale $\mu$.\footnote{This is in contrast to the one-loop running
  coupling $\lambda(\mu)$ which approaches a Landau singularity at
  high energies.} Perturbation theory, however, ceases to be
meaningful before reaching the fixed point.  In addition, lattice
calculations indicate that such large values for the coupling are
inconsistent with the requirement of small cutoff effects~\cite{LW}.
To accomodate these results we define two different cutoff conditions
for the running coupling at the cutoff scale $\Lambda$:
\begin{equation}
\lambda_c(\Lambda)=\lamFP/4 \quad\qquad {\rm and} \quad\qquad 
\lambda_c(\Lambda)=\lamFP/2\,.
\end{equation}  
The first choice corresponds to a running coupling which is definitely
perturbative at scale $\Lambda$ \cite{RW}, giving rise to a modest
25\% two-loop correction to the one-loop beta function $\beta_\lambda$
of the Higgs quartic coupling.  This condition gives $M_H$ upper
bounds below which the SM is certainly well-defined and perturbative.
The second choice, $\lambda_c(\Lambda)=\lamFP/2$, causes a 50\%
two-loop correction to $\beta_\lambda$, and its value is comparable
with upper bounds on $\lambda(\Lambda)$ which can be obtained from
lattice calculations~\cite{LW}.  It is also close to the upper bound
of the perturbative regime~\cite{RW}.  Hence these two choices for
$\lambda_c(\Lambda)$ summarize the uncertainty connected to the
precise formulation of the cutoff condition.  Carrying out a numerical
analysis \cite{HR} we find the uncertainty of the $M_H$ upper bound
related to the cutoff condition to be about $\pm$ 50 GeV for TeV
cutoff scales, and less than $\pm$ 3 GeV for cutoff scales
approaching $\Lambda_{Pl} \simeq 10^{19}$~GeV.

\begin{figure}[htb]
\vspace*{20pt}
\centerline{
\epsfysize=2.55in \rotate[l]{\epsffile{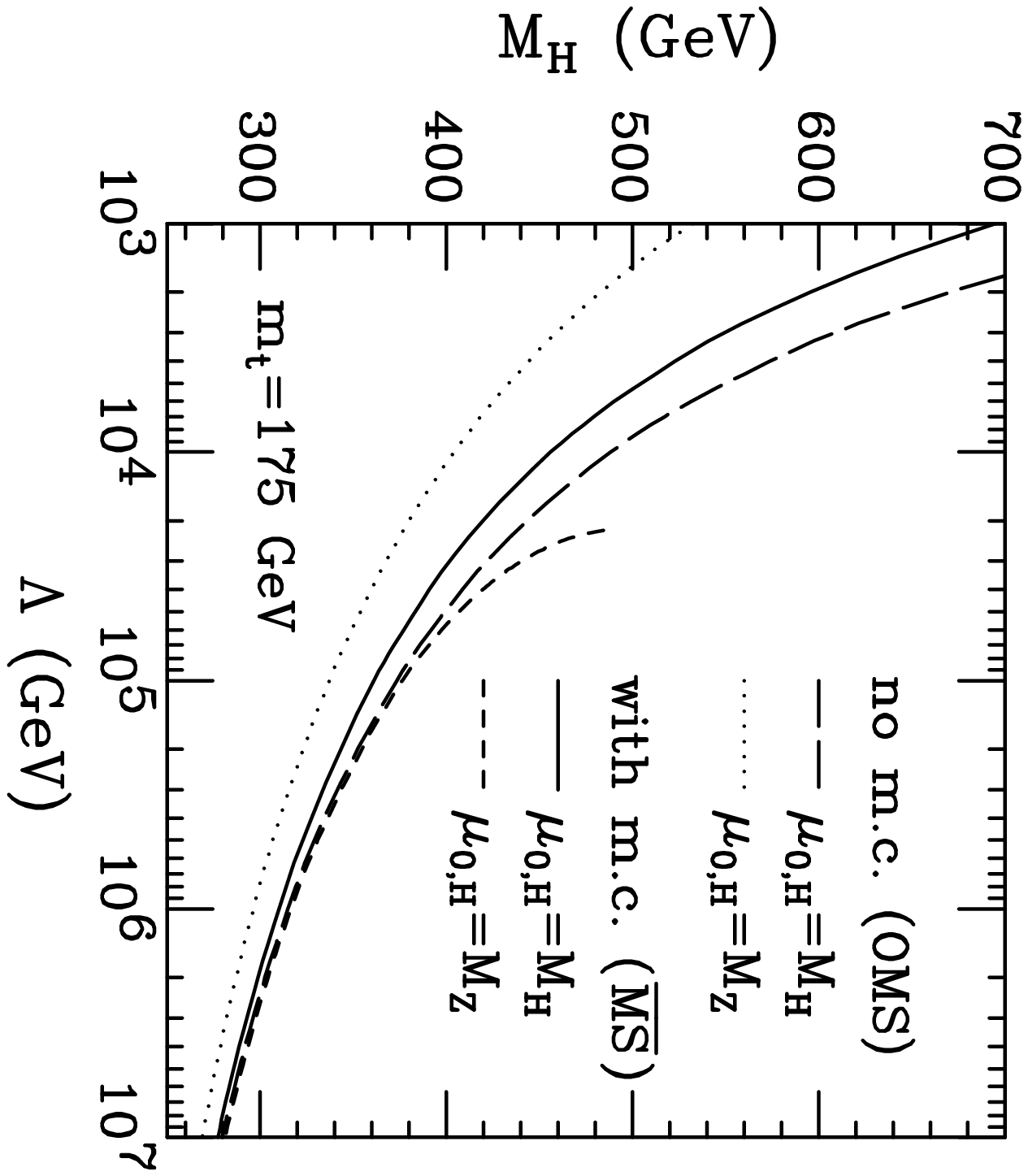}}
\hspace{0.8cm}
\epsfysize=2.5in \rotate[l]{\epsffile{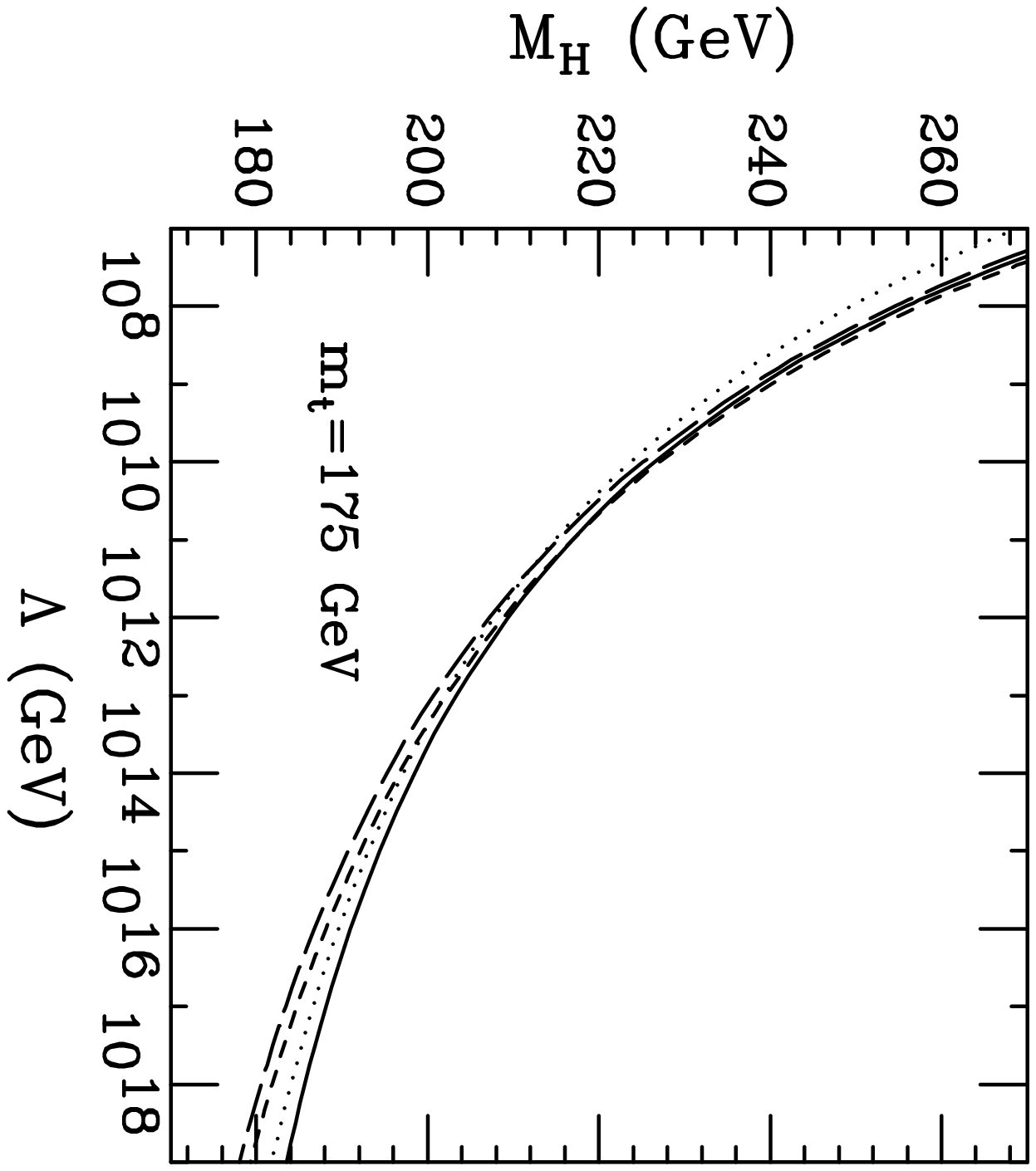}}
}
\small
\caption{Choosing two-loop RG evolution and cutoff condition
  $\lambda_c(\Lambda)=\lamFP/2$, the upper bound on $M_H$ is
  calculated. The running Higgs and Yukawa couplings, $\lambda(\mu)$
  and $g_t(\mu)$, are fixed by the physical masses $M_H$ and $m_t$
  using matching conditions with or without one-loop matching
  corrections. In addition, the Higgs matching scale is varied to be
  $\mu_{0,H}=M_H$ or $M_Z$.  The top-quark mass is fixed at $m_t=175$
  GeV, and $\mu_{0,t}=m_t$.  } \normalsize
\label{figmudep}
\end{figure}

A far more important point is the choice of the matching scale $\mu_0$
used in the definition of the couplings, Eqs.~(\ref{mchiggs}) and
(\ref{mctop}).  On first sight, it seems to be convenient to take
$\mu_0=M_Z$ since the gauge couplings have also been defined at this
scale.  This choice, however, introduces large corrections in the
matching correction $\delta_H$ if the Higgs mass is larger than
$\approx$ 400 GeV, and it results in the strange effect that the Higgs
coupling is restricted to a maximal value of 1.2 which is obtained for
$M_H=495$ GeV~\cite{HR}. This point can be identified as low-scale
end-point of the short-dashed curve in Fig.~\ref{figmudep}.  
Clearly the choice $\mu_0=M_Z$ is inappropriate for perturbative
calculations involving large values of $M_H$.  Neglecting the matching
conditions (setting $\delta_i=0, i=H,t$), the choice $\mu_0=M_Z$ also
leads to unreliable results, now resulting in a too stringent bound on
$M_H$ for low cutoff scales (dotted curve in Fig.~\ref{figmudep}).
This confirms that the appropriate choices for the Higgs and top quark
matching scales are $\mu_{0,H} \simeq$ max$\lbrace m_t,M_H \rbrace$
and $\mu_{0,t} \simeq$ $m_t$, respectively, as described in~\cite{HR}.

Using the choice $\mu_{0,H}=M_H$, we can also estimate the uncertainty
in the Higgs mass upper bound due to higher order corrections to the
matching corrections $\delta_H$ and $\delta_t$. This is done by
comparing the solid line (which includes matching corrections) and the
long-dashed line (without matching corrections) in
Fig.~\ref{figmudep}.  We find that the difference of the two results
exceeds 100 GeV at small embedding scale $\Lambda$, but reduces to
less than about 6 GeV at large scale.

\begin{figure}[htb]
\vspace*{20pt}
\centerline{
\epsfysize=3.8in \rotate[l]{\epsffile{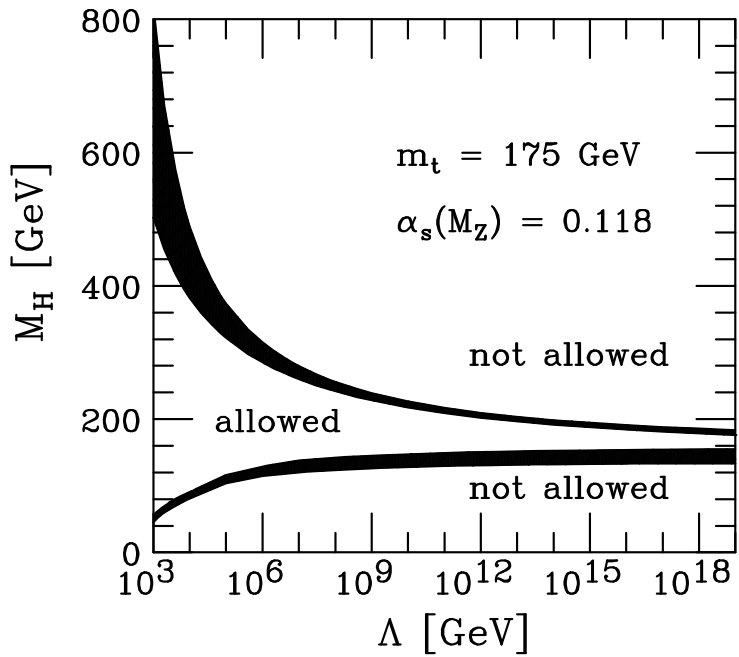}}
}
\small
\caption{Summary of the uncertainties connected to the bounds on $M_H$.
  The upper solid area indicates the sum of theoretical uncertainties
  in the $M_H$ upper bound for $m_t=175$ GeV~\protect\cite{HR}. The
  upper edge corresponds to Higgs masses for which the SM Higgs sector
  ceases to be meaningful at scale $\Lambda$ (see text), and the lower
  edge indicates a value of $M_H$ for which perturbation theory is
  certainly expected to be reliable at scale $\Lambda$.  The lower
  solid area represents the theoretical uncertaintites in the $M_H$
  lower bounds derived from stability requirements
  \protect\cite{altisi,casas1,casas2} using $m_t=175$ GeV and
  $\alpha_s=0.118$.} \normalsize
\label{figfinal}
\end{figure}

The sum of all theoretical uncertainties in the Higgs upper bound is
summarized in Fig.~\ref{figfinal}, taking $m_t=175$ GeV.  
In that figure we also show the results for the lower bounds using the
results from~\cite{altisi,casas1,casas2}, taking $m_t=175$ GeV and
$\alpha_s(M_Z)=0.118$.  We use two-loop beta functions and
appropriately choose the matching scale to be $\mu_{0,H}=M_H$.  We
vary the cutoff condition between $\lambda_c(\Lambda)=\lamFP/4$ and
$\lamFP/2$ as discussed above, and we carry out numerical calculations
both with and without inclusion of one-loop matching corrections
$\delta_H$ and $\delta_t$. For low cutoff scales, the $M_H$ lower
bound and its uncertainties are taken from~\cite{casas2}, and for
larger cutoff scales we use the condition $\lambda_c(\Lambda)=0$
\cite{altisi}.  Hence we are able to identify an ``allowed'' area for
which we find a well-behaved, perturbative and stable Higgs sector of
the SM, and we can identify ``disallowed'' areas for which the SM
Higgs sector is an inconsistent or unstable theory.  The black area
identifies the transition zone between these regions, indicating the
uncertainties related to the choice of the various cutoff criteria and
their perturbative implementation.

Looking at Fig.~\ref{figfinal} we conclude that a SM Higgs mass in the
range of $160$ to $170$ GeV results in a SM renormalisation-group
behavior which is perturbative and well-behaved up to the Planck scale
$\Lambda_{Pl} \simeq 10^{19}$ GeV.

The remaining experimental uncertainty due to the top quark mass is
not represented here and can be found in \cite{altisi,casas1,casas2}
and \cite{HR} for lower and upper bound, respectively.  In particular,
the result $m_t=175\pm6$ GeV leads to an upper bound
\begin{equation}
M_H < 180 \pm 4\pm 5 \;{\rm GeV} \qquad 
{\rm if} \qquad \Lambda=10^{19}\; {\rm GeV}, 
\end{equation}
the first error indicating the theoretical uncertainty, the second
error reflecting the residual $m_t$ dependence~\cite{HR}.

\end{document}